\documentclass[a4paper,aps,prb,twocolumn,floatfix,showpacs]{revtex4}
\usepackage{amssymb}
\usepackage{amsmath}
\usepackage[dvips]{graphicx}
\usepackage{bm}

\begin{document}

\title{Qualitative breakdown of the non-crossing approximation for the symmetric one-channel Anderson impurity model at all temperatures}


\author{C. N. Sposetti$\sp{1}$}
\author{L. O. Manuel$\sp{1}$}
\author{P. Roura-Bas$\sp{2,3}$}

\affiliation{$\sp{1}$Instituto de F\'isica Rosario (CONICET) and Universidad Nacional de Rosario, Rosario, Argentina;}
\affiliation{$\sp{2}$Consejo Nacional de Investigaciones Cient\'{\i}ficas y T\'ecnicas (CONICET) Argentina;}
\affiliation{$\sp{3}$Depto de F\'{\i}sica CAC-CNEA, Argentina.}

\begin{abstract}
The Anderson impurity model is studied by means of the self-consistent hybridization 
expansions in its non-crossing (NCA) and one-crossing (OCA) approximations. 
We have found that for the one-channel spin-$1/2$ particle-hole symmetric Anderson model, the NCA 
results are qualitatively wrong for any temperature, even when the approximation  
gives the exact threshold exponents of the ionic states. 
Actually, the NCA solution describes an overscreened Kondo effect, because it is the same
as for the two-channel infinite-$U$ single level Anderson model.
We explicitly show that the NCA is unable to distinguish between these two very different physical systems, 
independently of temperature. 
Using the impurity entropy as an example, we show that the low temperature
values of the NCA entropy for the symmetric case yield the limit $S_{imp}(T=0)\rightarrow \ln\sqrt{2},$ 
which corresponds to the zero temperature entropy of the overscreened Kondo model.   
Similar pathologies are predicted for any other thermodynamic property. 
On the other hand, we have found that the OCA approach lifts the artificial mapping between the models and 
restores correct properties of the ground-state,
for instance, a vanishing entropy at low enough temperatures $S_{imp}(T=0)\rightarrow0$.
Our results indicate that the very well known NCA should be used with caution close to the symmetric point of the Anderson model.
\end{abstract}

\maketitle

\section{Introduction}
Exact solutions of strongly correlated Hamiltonians are very specific and/or computationally expensive, 
therefore, it is often necessary to resort to approximate solutions. In order to provide a qualitative 
understanding, the approximations should recover basic physical features of the models. 
One of the most studied correlated Hamiltonian is the Anderson impurity model (AIM), 
originally proposed for the description of magnetic impurities in a conducting host \cite{anderson61}, 
which manifests the Kondo phenomenon at low enough temperatures \cite{kondo64}. 
Within the approximated schemes, the non- and one-crossing approximations  (NCA, OCA) have 
an important place in the literature \cite{nca, nca-names, pruschke89}. 
Its uses are not only restricted to impurity models \cite{jacob09},  
but also have been extended to correlated lattice models in the context of the dynamical mean field theory (DMFT) 
\cite{kotliar06}.

The scopes and limitations of both approximations are well known through detailed analysis of the 
corresponding semi-analytical expressions of the Green's functions, and also by comparison with exact techniques, 
like the Bethe ansatz and the numerical renormalization group calculations \cite{kim97,kim95,costi94}. 
In particular, a lot of effort has been dedicated to the analysis of the Fermi liquid properties of the AIM 
when it is solved within NCA and OCA; their successes and failures are quite well known 
and we refer the reader to the specific references \cite{pruschke89,haule01,tosi11}.
%
%
In brief, it has been argued that one of the underlying reasons for the incorrect description of the 
ground states properties within these approximations is their wrong predictions of the threshold exponents of the 
ionic states \cite{kroha98}. 

In this paper, we show that the NCA solution of the one-channel spin-$1/2$
particle-hole symmetric Anderson model yields the exact ionic threshold exponents at zero temperature.
Nevertheless, its results are qualitatively wrong at any 
temperature:  this solution physically corresponds to the two-channel infinite-$U$ single level 
Anderson model and, explicitly, we show that the NCA is unable to distinguish between the two 
physical systems. 
In other words, we conclude that the correct threshold exponents are a necessary but not sufficient 
condition for Fermi liquid behavior.

From a numerical analysis of the impurity contribution to the entropy, we exemplify that the NCA solution 
does not correspond to the model at hand. Surprisingly, the obtained solution is the one that arises 
from an overcompensated two-channel (2CH) spin-$1/2$ Anderson model. 
In fact, we obtain a finite residual entropy at very low temperatures, well below the Kondo one $T_K,$
being the extrapolated ground state value $S_{imp}(T=0) =  ln\sqrt{2},$ instead of zero expected 
for a Fermi liquid ground state. 
This particular fractional value of the ground state degeneracy is known to correspond to 
the 2CH Kondo model \cite{2ch}. 
%
%

Although we illustrate the failure through the impurity entropy, we also provide clear evidences of how this 
qualitative wrong solution extends to any other thermodynamic property, at any temperature. 
Furthermore, we prove that the NCA results for the electrical conductance \cite{gerace02}, when the symmetric AIM is
used to analyze transport properties through quantum dots, is the same as the expected conductance 
for the 2CH Kondo model. 

On the other hand, we show that the next leading order in the self-consistent hybridization expansion, 
the OCA scheme, lifts the artificial mapping between the models at the symmetric point, and gives a 
qualitative good prediction of the properties for the models at hand, with satisfactory quantitative 
improvements for large degeneracy of the impurity states.

Therefore, the present work is a warning for the potential users of such a simple approximation like NCA, that
can incorrectly map a given model into another. It is worth mentioning that, due to its simplicity, 
the NCA in its finite-$U$ version is nowadays widely used \cite{takemura15,oguri15}. 

\section{Model and formalism}
\label{modelo}

In this work, we consider the orbitally degenerate Anderson impurity model, 
represented by the following Hamiltonian:
\begin{eqnarray}\label{anderson}\begin{split}
H=& \sum_{km}\epsilon_{km} n_{km} + \sum_{m}E_{f} n_{m} + 
U \sum_{m>m'} n_{m}n_{m'}+ \\
& +\sum_{km}(V_{km} f^{\dagger}_{m} c_{km}+ {\rm H.c.}) ,
\end{split}\end{eqnarray}
where $m=m_j$ is the magnetic quantum number labeling the total angular momentum sector $j$,
with degeneracy $N=2j+1$. This representation is specially suitable for describing rare-earth 
systems, like Ce or Yb compounds, in which there is a strong spin-orbit coupling. 

In this Hamiltonian, the parameters $\epsilon_{km}$ ($E_{f}$) represent the energy of the conduction (impurity) 
electrons and $c^{\dagger}_{km}$ ($f^{\dagger}_{m}$) and $n_{km}$ ($n_{m}$) the creation and 
number operators, respectively. $U$ denotes the energy cost when the impurity is doubly 
occupied and we consider that this energy is the same for all the pairs of localized orbitals. 
The energies $V_{km}$  are the hybridization matrix elements between the impurity and delocalized 
electrons. 
The coupling of the impurity with the conduction band is encoded in the so-called hybridization 
function 
$\Delta_{m}(\omega)\equiv\pi\sum_{k} V_{km}^{2}\delta(\omega -\epsilon_{km})$. 

With respect to the impurity, we restrict its Hilbert space to the states 
without electrons (empty state), single occupied with only one electron in the 
orbital $m$, and doubly occupied with two electrons in the orbitals $m$ and $m',$ defining 
the atomic configurations $\{\vert 0 >,\vert m >, \vert mm'>\},$ respectively.  
The number of single and doubly occupied states are $N_s=N$ and $N_d=N(N-1)/2,$ respectively.
Therefore, the size of the impurity Hilbert space is given by $g = 1 + N_s + N_d$, which can be 
accessed through the evaluation of the impurity contribution to the entropy, at high enough temperatures.

Using a slave-boson and pseudo-fermion representation of the impurity states \cite{nca}, $\vert  \nu > =
\nu^\dagger \vert vac >$, being $\vert vac >$ the vacuum state without any impurity degree of freedom,
the physical impurity operator can be expressed by 
$f^{\dagger}_{m}=s^{\dagger}_{m}b+\sum_{m'\neq m}d^{\dagger}_{m'm}s_{m'}$. 
Here, the operators $b^{\dagger}, s^{\dagger}_{m}$, and $d^{\dagger}_{m'm}$ create over the vacuum the 
impurity states $\{\vert 0>, \vert m>, \vert mm'>\},$ respectively.

Employing this notation, the Hamiltonian (\ref{anderson}) reads
\begin{eqnarray}\label{anderson-auxiliary}\begin{split}
H=& \sum_{km}\epsilon_{km} n_{km} + \sum_{m}E_{f} s^{\dagger}_{m}s_{m}+\\
& +(2E_{f}+U) \sum_{m>m'}d^{\dagger}_{mm'}d_{mm'}+\\
& +\sum_{km}\left(V_{km} s^{\dagger}_{m}b \;c_{km}+ {\rm H.c.}\right)+\\ 
& +\sum_{kmm'}\left(V_{km} d^{\dagger}_{m'm}s_{m'}c_{km}+ {\rm H.c.}\right). 
\end{split}\end{eqnarray}

An approximate solution of the model can be obtained from a self-consistent perturbation expansion
in the hybridization hoppings $V$. The method leads to the solution of the following system for 
the approximated self-energies, at the OCA level \cite{tosi11}, 
\begin{eqnarray}\label{sistema_oca}
\begin{split}
 &\Sigma_b(\omega)=\int_{-\infty}^{\infty} \frac{d\epsilon}{\pi}f(\epsilon)\sum_{m}\Delta_{m}(\epsilon)G_{s_m}(\epsilon+\omega)
 \Lambda_{m}^{(0)}(\omega,\epsilon),\\
 &\Sigma_{s_m}(\omega)=\int_{-\infty}^{\infty} \frac{d\epsilon}{\pi}f(\epsilon)\left[\Delta_{m}(-\epsilon)G_{b}(\epsilon+\omega)
 \Lambda_{m}^{(0)}(\epsilon+\omega,-\epsilon)+\right.\\
 &~~~~~~~~~~ +\sum_{m'\neq m} \left.\Delta_{m'}(\epsilon) G_{d_{mm'}}(\epsilon+\omega)\Lambda_{mm'}^{(2)}(\epsilon+\omega,\epsilon)\right],\\
 &\Sigma_{d_{mm'}}(\omega)=\int_{-\infty}^{\infty} \frac{d\epsilon}{\pi}f(\epsilon) \left[ \Delta_{m}(-\epsilon)G_{s_{m'}}(\epsilon+\omega)
      \Lambda_{mm'}^{(2)}(\omega,-\epsilon) + \right. \\
 &~~~~~~~~~~~~~~~~~~~ + \left. \Delta_{m'}(-\epsilon)G_{s_m}(\epsilon+\omega)\Lambda_{m'm}^{(2)}(\omega,-\epsilon)\right],\\
 \end{split}
\end{eqnarray}
where the Green's functions $G_i$ have the usual Dyson expression, $G_i(z)=[z-\epsilon_i-\Sigma_i(z)]^{-1},$ 
and $f(\omega)$ is the Fermi function. 

The functions $\Lambda_{m}^{(0)}(\omega,\omega')$ and $\Lambda_{mm'}^{(2)}(\omega,\omega')$ 
represent vertex corrections to the self-energies, and incorporate all the diagrams containing only one 
crossing between conduction propagators \cite{pruschke89,haule01}. They are given by
\begin{eqnarray}\label{vertices_oca}
\begin{split}
&\Lambda_{m}^{(0)}(\omega,\omega')=~~~~~~1~~~~~~+\\
&\int_{-\infty}^{\infty}\frac{d\epsilon}{\pi}f(\epsilon)\sum_{m'\neq m}\Delta_{m'}(\epsilon)G_{s_{m'}}(\omega+\epsilon)
G_{d_{mm'}}(\omega+\omega'+\epsilon),\\
&\Lambda_{mm'}^{(2)}(\omega,\omega')=~~~~~~1~~~~~~+\\
&\int_{-\infty}^{\infty}\frac{d\epsilon}{\pi}f(-\epsilon)\Delta_{m}(\epsilon)G_{s_{m'}}(\omega-\epsilon)G_{b}(\omega-\omega'-\epsilon).
\end{split}
\end{eqnarray}

Neglecting the second term in the right-hand sides of Eqs.(\ref{vertices_oca}) 
corresponds to consider the self-energies at the NCA level in Eqs.(\ref{sistema_oca}),
which only contain a self-consistent summation of dressed diagrams of order $V^2$. 

Once the Green's functions are calculated self-consistently, the impurity contribution to a 
given thermodynamic quantity can be obtained from the partition function 
\begin{equation*}\label{zf}
Z_{f}(T) = \int_{-\infty}^{\infty} d\epsilon~e^{-\beta\epsilon}\left[\rho_b(\epsilon) + \sum_{m}\rho_{s_m}(\epsilon)
                           +\sum_{m'> m}\rho_{d_{mm'}}(\epsilon)\right],   
\end{equation*}
in which the spectral functions are related with the Green ones via 
$\rho_i(\omega)=-\frac{1}{\pi}\mathcal{I}m[G_i(\omega)]$.

\section{Numerical Results}\label{results}

For the numerical evaluation of the self-energies and Green's functions, we have employed a square 
hybridization of intensity $\Delta$ with a half-bandwidth $D$, which is related with the hopping 
$V$ via $\Delta=\pi V^2 / 2D$. Furthermore, we chose $\Delta=1$ as our unit of energy and 
$E_f=-4$, $D=10\Delta$. 

For the computation of the impurity entropy $S_{imp}(T)$, we have followed 
the approach given by Hettler \textit{et. al} \cite{hettler2}, instead of the standard derivation of the 
partition function, $S_{imp}(T)=-\frac{\partial \Omega_f}{\partial T},$ with $\Omega_f=-T ln(Z_f)$. The 
main steps of this procedure are discussed in the Appendix.
\\

\textbf{The infinite Coulomb repulsion limit}\\

In the case of an infinite Coulomb repulsion, the OCA scheme tends to the NCA one because 
the vertex function $\Lambda_{m}^{(0)}(\omega,\omega')$ goes to 1, while the double-occupied states are not 
taken into account. 

The large $N$ limit of the model Hamiltonian Eq. (\ref{anderson}) for $U\rightarrow\infty$, 
was extensively studied within the NCA approach to heavy fermions compounds. 
For details, we refer the reader to the appropriate 
references \cite{nca,kim95,kim97}. Here, we briefly review the results of the NCA entropy as a 
function of temperature for several values of the degeneracy $N$. This will be useful for the 
analysis of this property in the following sections. 

\begin{figure}[h!]
\includegraphics[clip,width=7.0cm]{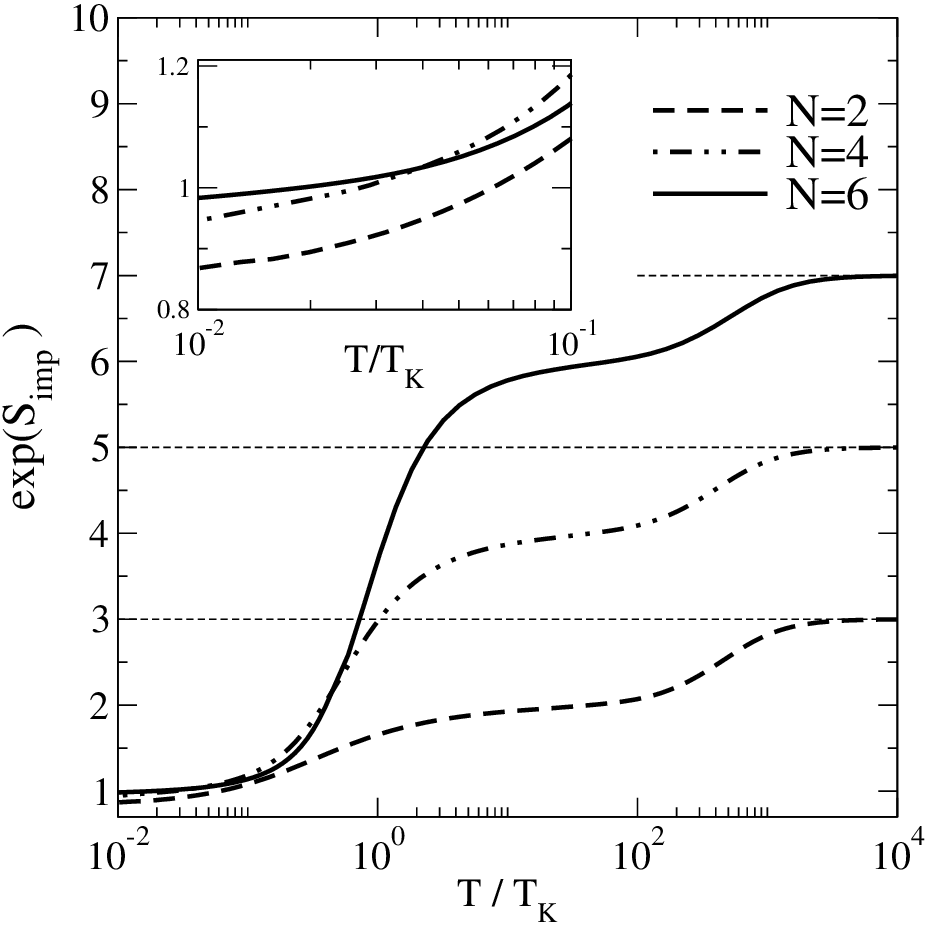}
\caption{Impurity contribution to the total entropy as a function of temperature for several values 
of the degeneracy $N$ and $E_f=-4$. The temperatures are scaled by the corresponding Kondo ones, being 
$T_K / \Delta \approx 0.007, 0.016,$ and $0.1$ for $N = 2, 4,$ and $6,$ respectively. 
The horizontal dashed line stands for a guide indicating the value of $g=1+N$. The inset shows the low 
temperature behavior.}
\label{fig-1.eps}
\end{figure}

In Fig. (\ref{fig-1.eps}) we present the impurity entropy as a function of temperature, for different values of the
degeneracy $N$. 
The temperatures are scaled by the corresponding Kondo ones, which we have extracted 
from the full width at half maximum (FWHM) of the spectral density (not shown), in good agreement with the 
exponential dependence of the parameters of the well known expression 
$T_K = D(N\Delta/D)^{1/N}\exp[\pi E_f / (N\Delta)]$ \cite{kroha98}. 
At large enough temperatures, $e^{S_{imp}}$ saturates at the value imposed by the local Hilbert 
space dimension $g=1+N$. As the temperature 
is lowered, an intermediate plateau can be observed in which $e^{S_{imp}} \simeq  N,$ due to the fact that 
now the impurity is almost single occupied (the empty state does not contribute to the entropy). 
When the temperature falls under $T_K,$ the value of $e^{S_{imp}}$ tends to one, 
as expected for the non-degenerate Kondo ground state. However, we can observe that for $N=2$, the entropy 
at very low temperatures, $T\ll T_K,$ becomes negative ($e^{S_{imp}} < 1$) as it is shown in the inset of 
Fig. (\ref{fig-1.eps}), pointing out the breakdown of the 
Fermi liquid properties. This shortcoming of the NCA is remedied as $N$ increases. 
It is a well known result and it is expected for a large-$N$ theory which 
becomes exact for $N\rightarrow\infty$. 
In fact, such deficiency is not longer found in the case of $N=6,$ for temperatures down to $T\sim 0.01 T_K,$ as 
has been noted previously \cite{otsuki06}.\\

\textbf{The NCA with finite Coulomb repulsion}\\

For finite values of the Coulomb repulsion $U$, the NCA consists in the approximation of 
both vertex corrections by $\Lambda_{m}^{(0)}(\omega,\omega')=\Lambda_{mm'}^{(2)}(\omega,\omega')=1$.
For a two-fold degenerate model ($N=2$), the set of self-energies in Eq. (\ref{sistema_oca}), at the NCA level,
takes the form
\begin{eqnarray}\label{sistema_nca}
\begin{split}
&\Sigma_b(\omega)=\frac{2\Delta}{\pi}\int_{-D}^{D} d\epsilon \;f(\epsilon)G_s(\epsilon+\omega),\\
&\Sigma_s(\omega)=\frac{\Delta}{\pi} \int_{-D}^{D} d\epsilon \;f(\epsilon)[G_b(\epsilon+\omega)+G_d(\epsilon+\omega)],\\
&\Sigma_d(\omega)=\frac{2\Delta}{\pi}\int_{-D}^{D} d\epsilon \;f(\epsilon)G_s(\epsilon+\omega),\\
&Z_{f}(T) = \int_{-\infty}^{\infty} d\epsilon\;e^{-\beta\epsilon}[~\rho_b(\epsilon) + 2\rho_s(\epsilon)+\rho_d(\epsilon)~],\\
\end{split}
\end{eqnarray}
with $\epsilon_b=0$, $\epsilon_s=E_f,$ and $\epsilon_d=2E_f+U$.\\

We start this subsection raising the following point for the two-fold degenerate
model: as we shall see, the NCA at the symmetric particle-hole case of the Hamiltonian, $\epsilon_d = 0$, exactly 
gives the threshold exponents of the auxiliary Green's functions at zero temperature. Therefore, which is 
the reason of the well known failing of the NCA description of the Fermi liquid properties in this case? 

In order to shed light over this point, here we calculate the threshold exponents and analyze them for the particle-hole 
symmetric case.
The auxiliary Green's functions at zero temperature, displays a power-law divergent behavior, $G_i\sim|\omega-E_0|^{-\alpha_i}$ with $i=b, s, d$,
in the limit $\vert \omega - E_0 \vert \ll T_0$ where $E_0$ represent the ground state energy of the model, below 
which $G_i$ are purely real and $T_0$ is an integration constant related with the Kondo temperature. 
The threshold exponents $\alpha_i$ are known exactly for the one channel problem \cite{haule01,kim97}, being 
all of them equal to $\alpha_i=1/2$ when $\epsilon_d = 0,$ due to the fact that the impurity occupation $n_{imp}=1$ in this case.

These exponents can be obtained within the NCA scheme by analyzing its zero temperature limit.
In this limit, Eqs. (\ref{sistema_nca}) transform in a set of differential 
equations that can be solved analytically in the limit of a large enough bandwidth of the conduction electrons, 
$D\rightarrow\infty$, and at the low frequency range, $\vert \omega - E_0 \vert \ll T_0$. A detailed analysis of this 
procedure can be found in the literature \cite{muller,kim97,cox93}, and will not be done here. 
However, here we present the solution of such system for the inverse Green's functions, $g_i(\omega)=-1/G_i(\omega)$,
\begin{eqnarray}\label{sistema_g}
\begin{split}
&g_s(\omega)=\frac{\pi T_0}{\Delta}\sqrt{g_b(\omega)g_d(\omega)},\\
&g_d(\omega)=g_b(\omega) + \epsilon_d,\\
&E_0 - \omega=\frac{\pi}{2\Delta} \int_0^{g_b} dg_b \;g_s,\\
&E_0 - \omega=\frac{\pi}{\Delta} \int_0^{g_s} dg_s \frac{g_b g_d}{\frac{\pi}{\Delta}g_b g_d + g_b + g_d},\\
\end{split}
\end{eqnarray}
with $T_0=\frac{\Delta}{\pi}\exp(\frac{\pi}{2\Delta}\frac{\Delta\epsilon_{sb}+\Delta\epsilon_{sd}}{2})$, where
$\Delta\epsilon_{ij}=\epsilon_i-\epsilon_j$.

Away from the symmetric point, $\epsilon_d>0$, the NCA exponents have been already calculated \cite{haule01,kim97}, 
$g_i(\omega) \approx \vert E_0 - \omega \vert^{\alpha_i}$ being $\alpha_b=2/3$ and $\alpha_s=1/3,$ while $g_d$ remains constant as 
$\omega\rightarrow E_0$.
However, in contrast to the previous case, the situation in which $\epsilon_d=0$ is quite different and has not been discussed in the literature.
Notice that in this case $g_d=g_b,$ and the system Eq. (\ref{sistema_g}) is simplified to
\begin{eqnarray}\label{sistema_g_simetrico}
\begin{split}
&g_s(\omega) = \frac{\pi T_0}{\Delta}g_b(\omega),\\
&E_0 - \omega = \frac{\pi}{2\Delta} \int_0^{g_b} dg_b ~ g_s,\\
&E_0 - \omega = \frac{\pi}{2\Delta} \int_0^{g_s} dg_s ~ g_b,\\
\end{split}
\end{eqnarray}
from which we obtain the exact power law dependence of the inverse of the Green's functions,
\begin{eqnarray}\label{solucion_g_simetrico}
\begin{split}
&g_s(\Omega) \simeq 2T_0\vert\Omega\vert^{1/2},\\
&g_b(\Omega) \simeq g_d(\Omega)=2\Delta\vert\Omega\vert^{1/2},\\
&\Omega=\frac{E_0 - \omega}{T_0}.
\end{split}
\end{eqnarray}

We can observe a discontinuity in the values of the exponents as a function of $\epsilon_d,$ at $T=0$. For finite temperatures 
our numerical results (not shown here) exhibit a crossover behavior, when $\epsilon_d$ approaches zero, between both set of threshold exponents.

Therefore, the NCA at the symmetric point gives the correct leading frequency dependence of the auxiliary Green's functions. 
Early, it was pointed out \cite{kroha98} that the NCA inadequacy in describing the Fermi 
liquid properties was related with the incorrect NCA threshold exponents. 
However, our results seems to oppose this statement. That is, even with the exact threshold exponents, the NCA at the symmetric
point still fails in describing the exact ground state, which is indeed a very peculiar result.

In the following, we clarify this apparent contradiction with the important result that correct threshold exponents are a necessary 
but not sufficient condition for Fermi liquid behavior.
In particular we discuss the impurity entropy as a function of temperature 
in the case of finite Coulomb repulsion $U,$ within the NCA approach. 
Its low temperature asymptotic behavior allows us to elucidate this controversy. 

\begin{figure}[h!]
\includegraphics[clip,width=7.0cm]{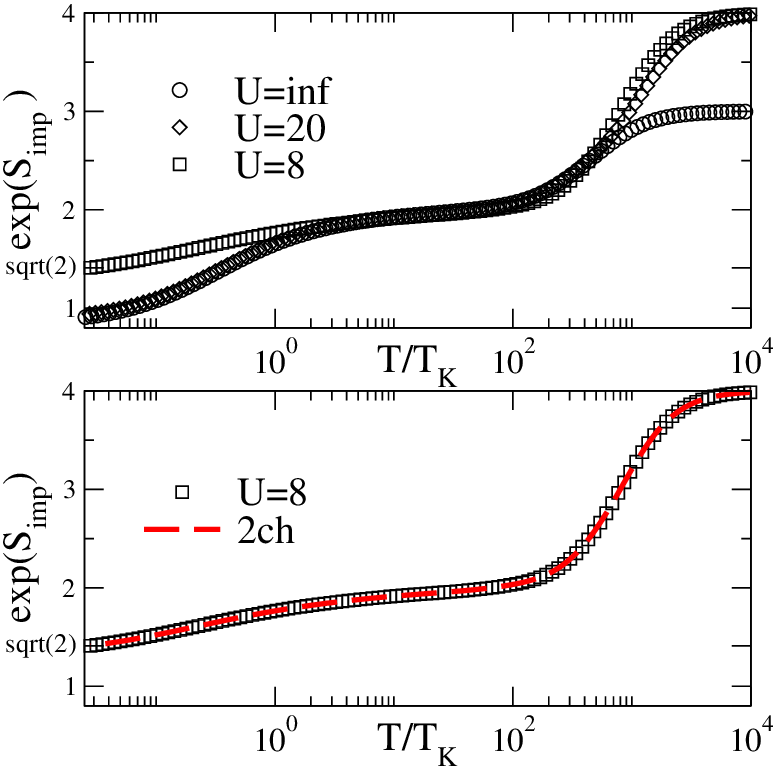}
\caption{(Color online) Upper panel: NCA impurity contribution to the total entropy as a function of temperature, 
for several values of the Coulomb repulsion and $N=2$ and $E_f=-4$. 
The temperatures are scaled by the corresponding Kondo ones, being 
$T_K / \Delta \approx 0.007, 0.0065,$ and $0.004$ for $U = \infty, 20,$ and $8,$ respectively. 
Lower panel: Same results as in the upper panel for $U=8$ and $E_f=-4$ and for the impurity entropy of the 
2CH, $U = \infty$ model at the same energy level  $E_f=-4$ (red dashed line).}
\label{fig-2.eps}
\end{figure}

In the upper panel of Fig. (\ref{fig-2.eps}) we show $S_{imp}$ for a two-fold degenerate
model ($N=2$) when the values of $U$ are 
lowered from $U=\infty$ to $U=8$, which corresponds to the symmetric case of the Hamiltonian ($E_f = -U/2$),
as a function of temperature in units of the Kondo one.

Note that the Kondo scale depends weakly on $U$ within NCA and it is strongly underestimated
for the symmetric case, given a value of $T_K\approx0.004$ for the present parameters, instead of 
$T_K\approx 0.086$ expected from the Haldane expression, 
$T_K=\sqrt{\frac{U\Delta}{2}}\exp(-\frac{\pi U}{8\Delta})$ \cite{haldane}.
While this is one of the well known quantitative shortcomings of the finite-$U$ NCA \cite{pruschke89,haule01,tosi11}, 
we find a qualitative wrong low temperature dependence of the thermodynamic properties
when the technique is applied to the symmetric case, even when, as we have shown, the auxiliary Green's functions have 
the correct low energy dependence.
In spite of the Kondo scale, in Fig. (\ref{fig-2.eps}) we show that away from the symmetric point, 
$e^{S_{imp}}\rightarrow1$ as the temperature is lowered, as expected. For any value of $U$ away from 
the symmetric point, the NCA entropy displays a behavior compatible with a low 
temperature non degenerate ground state.   
A qualitative deviation appears when the NCA is applied to the symmetric case: remarkably, we find in this case 
that the entropy remains finite at low temperatures, being its asymptotic behavior 
$e^{S_{imp}}\rightarrow\sqrt{2}$.
This deviation is not related with the previous mentioned underestimation of the  Kondo temperature, 
or the well-known violation of the Fermi liquid description, or with the recently highlighted 
NCA failure of the self-energies at high frequencies \cite{millis}.

Instead of that, we trace back this failure to an important artifact of the NCA: the approximation 
cannot be able to distinguish between a model with a $N$-degenerate ground state and $M$-excited
ones from another model in which $M$ conduction channels of spin $s=1/2$ are screening out an impurity 
with spin $S=(N-1)/2$. 
Some indications in this direction were pointed early, in the context of multichannel
Kondo models for heavy fermions compounds \cite{kim97}.

The two-channel (2CH) spin-1/2 Anderson model is described by the following Hamiltonian using the same auxiliary
particle representation \cite{hettler2}
\begin{eqnarray}\label{2ch-model}\begin{split}
H_{2CH} = &\sum_{k\sigma\tau} \epsilon_{k}n_{k\sigma\tau} + \sum_{\sigma}E_{f} s^{\dagger}_{\sigma}s_{\sigma}+\\
     &+ \sum_{k\sigma\tau}\left(V_{k\tau} s^{\dagger}_{\sigma}b_{\bar{\tau}}c_{k\sigma\tau} + {\rm H.c.}\right),
\end{split}
\end{eqnarray}
in which there are two independent conduction bands labeled by the index $\tau=1,2$, that transform according to
representations of the $SU(2)$ group. The large repulsion limit $U\rightarrow\infty$ is implicitly taken. 
Usually, the physical operator that creates 
an electron in the level $\sigma$ from a conduction electron in channel
$\tau$ is represented by $d^{\dagger}_{\sigma\tau}$ and the two boson flavors indicate an excited doublet of unoccupied local levels. 
In the Kondo regime limit, this model maps into the two-channel Kondo one, 
representing a single impurity of spin $s=1/2$ ($\sigma=\uparrow, \downarrow$) screened by two conduction bands. 
The difference between the Hamiltonian (\ref{2ch-model}) and the multiorbital one channel one, given by Eq. \ref{anderson-auxiliary} is 
evident. 
However, as we shall see, the NCA equations for the one channel symmetric AIM are actually identical to those 
appearing in the solution of the 2CH model. 

The lower panel of Fig. (\ref{fig-2.eps}) displays a comparison between the impurity entropy
for the symmetric case, $U=8,$ and the corresponding NCA one obtained from a 2CH, $U = \infty$ model 
(red dashed line) calculated from an independent code.
Both calculations produce identical results in the whole range of temperature. 
This confirm the previous statement.

Now we shall give an analytical demonstration of such coincidence.
Taking into account that, in the symmetric case, the energies $\epsilon_b=\epsilon_d=0$
become degenerate, the system of self-energies is simplified to
\begin{eqnarray}\label{sistema_nca_degenerado}
\begin{split}
&\Sigma_b(\omega)=\frac{2\Delta}{\pi}\int_{-D}^{D} d\epsilon f(\epsilon)G_{s}(\epsilon+\omega),\\
&\Sigma_s(\omega)=\frac{2\Delta}{\pi}\int_{-D}^{D} d\epsilon f(\epsilon)G_{b}(\epsilon+\omega),\\
&Z_{f}(T) = 2\int_{-\infty}^{\infty} d\epsilon~e^{-\beta\epsilon}[~\rho_b(\epsilon) + \rho_s(\epsilon)~].\\
\end{split}
\end{eqnarray}
The system in Eq. (\ref{sistema_nca_degenerado}) determines the whole set of thermodynamic 
properties of the model for a given temperature $T$. 
As it can be shown \cite{kim97}, this system of 
equations also arises as the NCA solution of the 2CH spin-$1/2$ Anderson model (with $U=\infty$),
for which a residual entropy $\ln\sqrt{2}$ was found not only within NCA 
\cite{kim97,solange2} but also from several exact techniques \cite{BA-3, parcollet} and the
numerical renormalization group \cite{solange-1}.

We want to stress here that the system in Eq. (\ref{sistema_nca_degenerado}), which we have
derived from the NCA solution of the one-channel spin-$1/2$ symmetric Anderson model, 
\textit{correctly} describes the non-Fermi liquid 
physics of the 2CH model, as it is corroborated in a comparison with  
exact Bethe ansatz and conformal field  theory (CFT) results \cite{kim97, cox93}. 
For instance, in addition to the residual entropy discussed above, a square root law dependence of the 
impurity resistivity of the 2CH model, $\rho(T)=\rho(0)[1-a\sqrt{T/T_K}]$, 
at low enough temperatures, was found using Eq. (\ref{sistema_nca_degenerado}), in very good agreement 
with the expected scaling dimension analysis from CFT \cite{kim97}. 

Note that the exact threshold exponents in the multichannel case $N\ge2$, $M\ge2$, obtained from CFT 
in the Kondo limit \cite{affleck-ludwig}, are given by $\alpha_b=N/(M+N)$ and $\alpha_s=M/(M+N)$, which are 
$1/2$ for $N=M=2$. 
The known NCA accuracy when applied to the multichannel $N=M=2$ case is related with 
its success in giving these exact exponents. 
Here we found the answer to our question made in the Introduction: 
the NCA fails when solving the one channel symmetric AIM, even with the correct threshold exponents, 
simply because it is solving another model, that is the $N=M=2$ multichannel one. 

It is worth to note that the equivalence, at the NCA level, of the two different models
is valid for any temperature $T$, as it is shown in the lower panel of Fig. (\ref{fig-2.eps}),
and for any frequency $\omega,$ as it is shown in Eq. (\ref{sistema_nca_degenerado}).
Not only the thermodynamic properties given by the NCA solution of the one channel symmetric AIM are 
qualitative wrong, but also some other observables.
This follows from the fact that, for instance, the physical spectral density (as any other correlation function) 
is built as a convolution of the auxiliary ones, and these spectral functions, obtained from Eq. 
(\ref{sistema_nca_degenerado}), are actually describing the 2CH model \cite{kroha98}.
As an important example, we consider in detail the electric conductance obtained from
the NCA solution of the one-channel $N=2$ symmetric AIM. Starting from the physical spectral density 
per spin $\sigma$,
\begin{eqnarray}\label{rho_nca_simetrica}
\begin{split}
&\rho_{f}(\omega)=\rho_{bs}(\omega)+\rho_{ds}(\omega),\\
&\rho_{bs}(\omega)=\int_{-\infty}^{\infty} \frac{d\epsilon}{Z_ff(-\omega)}~e^{-\beta\epsilon}\rho_b(\epsilon)
\rho_s(\epsilon+\omega),\\
&\rho_{ds}(\omega)=\int_{-\infty}^{\infty} \frac{d\epsilon}{Z_ff(\omega)}~e^{-\beta\epsilon}\rho_d(\epsilon)
\rho_s(\epsilon-\omega),\\
\end{split}
\end{eqnarray}
the equilibrium conductance $G(T)$ can be written as follows \cite{win}
\begin{eqnarray}\label{G1ch-simetrica}
\begin{split}
&G_{1CH}(T)=G_0 \times 2\pi\Delta\sum_{i}\int_{-D}^{D} d\epsilon (-f'(\epsilon))\rho_{is}(\epsilon) 
\end{split}
\end{eqnarray}
where the index $i$ labeled the two degenerate states, namely $i=b,d,$ and $G_0 = 2e^2/h$. 
Note that under the change  $\omega\rightarrow-\omega$ in the last expression of 
Eq. (\ref{rho_nca_simetrica}), $\rho_{ds}(-\omega)=\rho_{bs}(\omega),$ provided that the relation 
$\rho_b(\epsilon) = \rho_d(\epsilon)$ is fulfilled from Eq. (\ref{sistema_nca_degenerado}).
After a permutation in the order of the integrals, and a single change of variables in 
Eq. (\ref{G1ch-simetrica}), the two contributions to $G_{1CH}(T)$ become identical. 

The right-hand side of Eq. (\ref{G1ch-simetrica}), with identical 
$\rho_{is}(\epsilon)$ spectral functions (or changing $\omega\rightarrow-\omega$
if necessary), is also obtained for the conductance in Ref. \cite{win} when it is applied to the 
overscreened 2CH-model arising from the system in Eq. (\ref{sistema_nca_degenerado}) \cite{solange2}. 
In this case, the index $i$ represents the two different bath of conduction electrons related by 
$SU(2)$ symmetry that we have denoted by the index $\tau$ in Eq. (\ref{2ch-model}). 
Therefore, we conclude that the NCA conductance for the two physically 
different models, and for a given temperature $T,$ are the same, 
\begin{eqnarray}\label{G1ch-G2c}
\begin{split}
&G_{1CH}(T)=G_{2CH}(T).
\end{split}
\end{eqnarray}
We have verified the above relation numerically.

This result reinforces the artificial NCA mapping between the two physical systems. 
While a square root temperature dependence of the conductance ($G_{2CH}(T)\approx a -b\sqrt{T}$) 
at low enough temperature was found in the case of the overscreened Anderson model using several techniques \cite{solange-1,
hettler, hettler2},
this is not the expected behavior of the one-channel spin-$1/2$ Anderson impurity. 
The former follows a square  dependence ($G_{1CH}(T)\approx a +bT^2$) at low temperatures as it can be
seen from both, fitting experimental conductance measurements and 
numerical renormalization group calculations \cite{gold,G_E}.

The temperature dependence of the NCA $G_{1CH}(T)$ at the symmetric point of the model was 
numerically studied in Ref. \cite{tosi11}. In that work, specially in Fig. (5), a deviation
of the NCA $G_{1CH}(T)$ as compared with the exact one for all temperatures was found. 
As we previously mentioned, it is not expected that such a simple approximation can be satisfactorily 
compared with exact results due to the several deficiencies that it suffers, 
like the underestimation of the Kondo scale and the violations of Fermi liquid properties. 
However, what we are showing here is a deeper source of such deviation. 
What we state in this work is that such NCA conductance correctly describes an overscreened Kondo 
model instead of the ordinary one channel spin-$1/2$ symmetric model, 
which is the subject of that reference \cite{tosi11}.
In addition, transport properties that include the symmetric point of the AIM were studied by 
using NCA in Ref.\cite{gerace02}.
Based in the conclusion of the present paper, these results should be taken with caution.

It should be mention that the NCA physical spectral functions, in their frequency dependence, 
are trivially different for both models: for instance, in the symmetric one-channel AIM, the 
symmetry condition implies the existence of two charge fluctuation resonances, located symmetrically
around the Fermi level, at $\omega \sim \pm E_f$; while for the two-channel AIM, the infinite $U$ 
condition implies the existence of only the empty-simply occupied resonance at $\omega \sim E_f$. 
The NCA identity, Eq. (\ref{G1ch-G2c}), arises because the conductance is related with a frequency 
integral of the physical spectral functions.\\

\textbf{Inclusion of the crossing contributions}\\

In this subsection we analyze the inclusion of crossing diagrams to the auxiliary
self-energies by solving together Eq. (\ref{sistema_oca}) and Eq. (\ref{vertices_oca}).
It is expected that the one-crossing approximation lifts the equivalence, at the 
NCA level, of the symmetric one-channel and the two-channels overscreened models. 

From the self-consistent system in Eq. (\ref{sistema_oca}), it can be observed that
the inclusion of the crossing diagrams, which are of the order of $V^4$, explicitly  
introduces an asymmetry between this system and the corresponding one that emerge in 
the 2CH case, Eq. (\ref{sistema_nca_degenerado}).
Note that the usual 2CH spin-$1/2$ models involve infinite Coulomb repulsion and therefore, 
the next leading order in the self-consistent hybridization expansions is given by diagrams of
order of $V^6.$ \cite{nca}
Therefore it is expected 
that the NCA artificial mapping between two different physical models should be broken.

A detail study and comparison between the NCA and OCA solutions of the AIM can be found 
in the work of R\"{u}egg \textit{et al.} \cite{millis} and it is not our purpose to reproduce it here. 
In this paper we restrict ourselves to show  the rupture of the artificial equivalence
between the models when the vertex corrections in Eq.(\ref{vertices_oca}) are 
taking into account.

\begin{figure}[h!]
\includegraphics[clip,width=7.0cm]{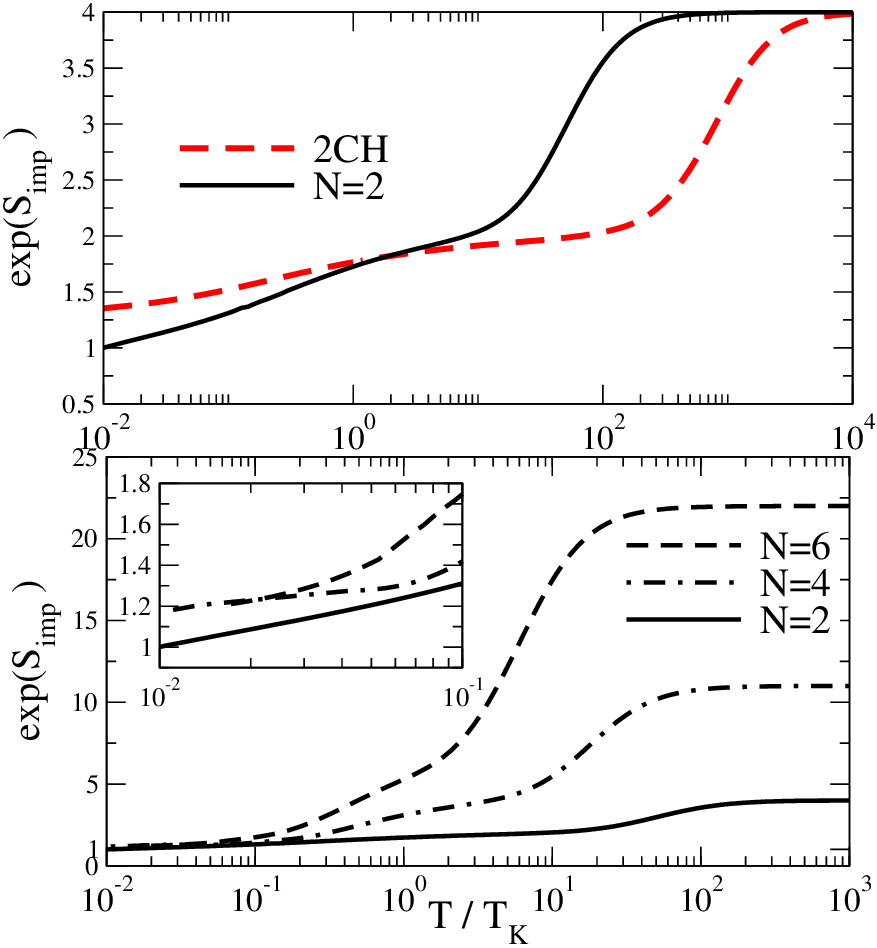}
\caption{(Color online) Upper panel: Comparison of the impurity entropy for the symmetric $N=2$ AIM 
calculated within OCA and the corresponding one for the overscreened model using Eq. (\ref{sistema_nca_degenerado}) 
at the same energy level  $E_f=-4$.
Lower panel: Impurity contribution to the total entropy as a function of temperature, 
within the OCA approximation, for several values 
of the degeneracy $N$ and for $U=8$, $E_f=-4$. 
The temperatures are scaled by the corresponding Kondo ones being 
$T_K / \Delta \approx 0.06, 0.185,$ and $0.6$ for $N = 2, 4,$ and $6,$ respectively.
The inset shows the low temperature behavior.}
\label{fig-3.eps}
\end{figure}

Once again, we present calculations for the impurity entropy for a 
finite value of $U$ and different values of $N$. These results are not shown in the work of 
R\"{u}egg \textit{et al.} \cite{millis}
In Fig. (\ref{fig-3.eps}) we plot the impurity entropy as a function of temperature using
the OCA approach for several values of the degeneracy $N$ and for $U=8$.

Note that the Kondo scale obtained from OCA for $N=2$ is largely improved ($T_K = 0.06$) 
with respect to the NCA one ($T_K = 0.004$), in relation with the Haldane value ($T_K = 0.086$). 

The upper panel of Fig. (\ref{fig-3.eps}) shows a different temperature dependence of the 
calculated OCA impurity entropy for the symmetric one-channel AIM as compared with the corresponding
one for the overcompensated model.  Furthermore, the low temperature values of the OCA 
calculation for the symmetric one-channel restores the expected value of the entropy. 
As in analogy to the case of the infinite-$U$ limit, while OCA restored the right tendency at 
low enough temperatures,
($S_{imp}\rightarrow 0$), it still is inadequate for a correct description of the Fermi liquid 
properties, note the tendency towards negative values of the entropy for $T\ll T_K$ shows in the 
inset of Fig. (\ref{fig-3.eps}, lower panel).
The lower panel of Fig. (\ref{fig-3.eps}) shows the OCA impurity entropy for the degenerate AIM for
several values of $N$. 
In contrast to the infinite $U$ limit, the large$-N$ results of the OCA approach have not being 
extensively studied. We do not find negative entropy in the cases of $N=4$ and $N=6$. 
Regarding the behavior of the NCA solution for finite $U$ as a function of $N$, we remembered to 
the reader that the qualitative behavior of the thermodynamic properties, in particular the entropy,  
are improve as $N$ is increased, however the Kondo temperature is still underestimated in comparison 
with the OCA one.

It is worth noticing that there is no correspondence between the threshold exponents, as obtained by 
NCA or by OCA, and the low temperature entropy as obtained by the same approximation. This is seen if 
one considers that, although OCA does not change the NCA values of the threshold exponents \cite{cox93}, 
it gives, as we have shown numerically, the correct low temperature entropy behaviour, as opposed 
to what happens with NCA.

\section{Summary} \label{conclusions}

We have used the non- and one-crossing self-consistent hybridization expansions
as approximate solvers for the $N$ degenerate Anderson impurity model.
After a brief review section of the entropy results given by the NCA in its infinite $U$
limit, we have focused in the results of the NCA solution of 
the particle-hole symmetric one-channel spin-$1/2$ Anderson Hamiltonian.
Our results show that, in this case, the exact threshold exponents of the auxiliary Green's functions
are recovered by the NCA. However, the later does not mean that the Fermi liquid properties are reproduced.
We have addresses this apparent contradiction showing that the correct threshold exponents are a necessary 
but not sufficient condition. We have showed that the system of 
self-consistent equations for the ionic 
self-energies and the partition function is the same as the one that arise from the 
overcompensated two-channel model (with $U=\infty$) and, therefore, the NCA 
cannot distinguish between these two very different Hamiltonians for any temperature.

Numerically, we have illustrated this NCA failure by means of the computation of the impurity entropy, 
which, in turns, exhibits a residual fractional entropy at very low temperatures, in agreement with 
the expected result for the two-channel Kondo model. 
Furthermore, we have proven that the electronic conductance through the impurity is the same that the 
corresponding one for the NCA conductance of the 2CH model. 
In addition, the lower (upper) charge-transfer peak for a given spin component, $\rho_{bs}(\omega)$ 
($\rho_{ds}(-\omega)$), is exactly the same that in the case of the overscreened model, per channel and
per spin $\rho_{si}(\omega)$. This means that any other  
observable --dynamic and thermodynamic-- that depends on the spectral density, 
will appear to be the same as for the 2CH Anderson model. 

This peculiar and very pathological result of the NCA is fixed when vertex corrections are introduced through the OCA impurity solver. 
Specifically for the symmetric point of the one-channel AIM, the known shape and tendency of the 
properties as a function of the model parameters are recovered by OCA.

The discussion presented here, the qualitative breakdown of NCA for the particle-hole symmetric AIM, 
becomes of general interest because, nowadays, NCA is one of the most used impurity solver methods, due to 
its simplicity and straightforward extension to more complex situations.
Therefore, our work highlights the importance of vertex corrections, even for a qualitative description 
of Anderson impurity models.

As a concluding remark, we mention that, probably,  artifacts as the one we have found in this work could also 
be found when solving correlated models by non-crossing re-summations of diagrams in the 
evaluation of self-energies. These approaches, known under the generic name of self-consistent 
second Born approximations, are widely used in the literature. In view of our results, 
a detail analysis of the inclusion of higher order diagrams may be done in order to test the reliability of the 
approximations.

\acknowledgments
This work was partially supported by PIP 00273, PIP 01060  and 
PIP 112-201101-00832 of CONICET, and PICT R1776  and PICT 2013-1045 
of the ANPCyT, Argentina. 

\section*{appendix}
\begin{center}
 \textit{Numerical evaluation of the NCA equations and entropy}\\
\end{center}
 
In this appendix we give details of the calculation we have implemented 
in order to get an accurate solution of the NCA equations at low temperatures, 
which is crucial for getting the lowest temperature values of the impurity
entropy. 
 
As we have mentioned at the beginning of the numerical results section, 
we have closely followed the approach given by Hettler, Kroha, and 
Hershfield in  Ref. \onlinecite{hettler2} for the computation of the 
impurity entropy $S_{imp}(T)$. However, we have found a better accuracy for the solution of 
the NCA selfconsistent equations, Eq. (\ref{sistema_oca}), through a different choice for the Lagrangian 
parameter, $\lambda_0$, as we will describe in what follows.

Note that the whole set of physical properties of the model Hamiltonian in Eq. (\ref{anderson-auxiliary}) 
does not depend on 
the individual values of the pseudo particle energy levels, $\epsilon_i$,  but depend on their differences. 
Consider, for instance, the  physical 
spectral function for transitions between the vacuum and single occupied states, showed in Eq. (\ref{rho_nca_simetrica})  

\begin{eqnarray}\label{rho_nca_bs}
\begin{split}
&\rho_{bs}(\omega)=\int \frac{d\epsilon}{Z_f~f(-\omega)}~e^{-\beta\epsilon}\rho_b(\epsilon)
\rho_s(\epsilon+\omega),\\
&Z_{f} = 2\int d\epsilon~e^{-\beta\epsilon}[~\rho_b(\epsilon) + \rho_s(\epsilon)~].\\
\end{split}
\end{eqnarray}

If one shifts the energies $\epsilon_i\rightarrow\epsilon_i-\lambda_0$ of the pseudo particles 
and, at the same time, performs variable change $\epsilon \rightarrow \epsilon-\lambda_0$, the system
in Eq. (\ref{rho_nca_bs}) remains the same.

This freedom in the definition of the pseudo particle energies derives, in fact, from a symmetry of the Hamiltonian
related with a gauge transformation of the auxiliary operators according to 
$ \hat{\nu_i} \rightarrow  e^{i\lambda_0 t}\hat{\nu_i}$. We refer the reader to the appropriated 
references for a deep discussion of this issue \cite{kroha98,hettler2}.

According to this shift, the generic partition function reads
\begin{equation}\label{z-lambda_0}
Z_{f} = e^{-\beta\lambda_0}~\int d\epsilon~e^{-\beta\epsilon}\sum_{i} ~\rho^{\lambda_0}_{i}(\epsilon),
\end{equation}
being $\rho^{\lambda_0}_{i}(\omega)$ the pseudo particles spectral densities given by
\begin{equation}\label{rho_i}
\rho^{\lambda_0}_{i}(\omega)= \frac{ Im \Sigma_i(\omega)}{(\omega+\lambda_0-\epsilon_i-Re\Sigma_i(\omega))^2 + 
                                                      (Im \Sigma_i(\omega))^2 }.
\end{equation}

The numerical trick used by Hettler and co-workers consists in the determination of $\lambda_0$, at each NCA 
iteration, in such a way that the magnitude $Q^{\lambda_0}(\beta)\equiv
\int d\epsilon~e^{-\beta\epsilon}\sum_{i} ~\rho^{\lambda_0}_{i}(\epsilon)=1$. With this important trick,
the narrow peaks developed by the auxiliary spectral densities at the threshold energy moves towards, as the 
temperature decreases, to the Fermi energy, usually fixed at $\omega=0$. This allows the use of a dense 
grid of energy points around the zero frequency, with the consequent increment in resolution of such narrow peaks.
The final parameter $\lambda_0$ has the interpretation of the impurity contribution to the free energy, 
$Z_{f} = e^{-\beta\lambda_0}=e^{-\beta F_{imp}(T)}$. \\

In our work, we have made used of a different choice for the parameter $\lambda_0$. At the end of each NCA/OCA 
iteration we compute the following energies
\begin{equation}
\lambda^{i}_{0}=\epsilon_i + Re\Sigma_i(\omega=0),
\end{equation}
and fix $\lambda_0=min~\{\lambda^{i}_{0}\}$. With this choice, the narrow peak corresponding to the lowest energy
pseudo particle (and also the coherent contribution of the excited ones) is \textit{exactly} located at the 
Fermi energy $\omega=0$ for the next iteration. 
In our experience, we found that the present algorithm is even better than the previous one used by Hettler
\textit{et al.}.
Furthemore, the speed and stability of the calculation is improved due to the fact that there is no search for 
roots of non linear equations like $Q^{\lambda_0}(\beta)=1$. In our case, the magnitude $Q^{\lambda_0}(\beta)$  
depends on temperature and therefore we cannot identify the parameter $\lambda_0$ as the free energy. Instead of
that, the free energy reads as follows
\begin{equation}
F_{imp}(T) = \lambda_{0}(T) -T~ ln(Q^{\lambda_0}(T)).
\end{equation}

Finally, we compute the impurity entropy as a numerical differentiation of the free energy, 
$S_{imp}(T)=-dF_{imp}(T)/dT$.

Apart from the preceding discussion, we have followed the remaining steps regarding the numerical treatment of the 
NCA equations as indicated by Hettler \textit{et al.} in Ref. \onlinecite{hettler2}.


\begin{thebibliography}{99}
\bibitem{nca-names} By the name NCA we mean to the approximation obtained by perform a self-consistent 
hybridization expansion in which the ionic self-energies are characterized by bubbles of second order in 
the hybridization strength independently of being $U$ finite or not.
\end{thebibliography}
\end{document}